%% file: parameter-fitting.tex
\documentclass[nonacm, sigconf]{acmart}

\usepackage[ruled,vlined,linesnumbered]{algorithm2e}

\newcommand\norm[1]{\lVert#1\rVert}
\newcommand{\E}[1]{\mathrm E\hspace{-0.2em}\left[#1\right]}
\newcommand*{\ER}{Erd\H{o}s--Rényi\xspace}

\usepackage[textsize=footnotesize]{todonotes}
\usepackage{subcaption}
\usepackage{siunitx}
\usepackage{booktabs}
\usepackage{multirow}
\usepackage{placeins}

\AtBeginDocument{%
  }

\begin{document}

\title{Robust Parameter Fitting to Realistic Network Models via Iterative Stochastic Approximation}

\author{Thomas Bläsius}
\affiliation{%
  \institution{Karlsruhe Institute of Technology}
  \city{Karlsruhe}
  \country{Germany}}
\email{thomas.blaesius@kit.edu}

\author{Sarel Cohen}
\affiliation{%
  \institution{The Academic College of Tel Aviv-Yaffo}
  \city{Tel Aviv}
  \country{Israel}}
\email{sarelco@mta.ac.il}

\author{Philipp Fischbeck}
\email{philipp.fischbeck@hpi.de}
\author{Tobias Friedrich}
\email{tobias.friedrich@hpi.de}
\affiliation{%
  \institution{Hasso Plattner Institute, University of Potsdam}
  \city{Potsdam}
  \country{Germany}
}

\author{Martin S. Krejca}
\affiliation{%
  \institution{LIX, CNRS, Ecole Polytechnique, Institut Polytechnique de Paris}
  \city{Palaiseau}
  \country{France}}
\email{ martin.krejca@polytechnique.edu}

\renewcommand{\shortauthors}{Bläsius et al.}

\begin{abstract}
  Random graph models are widely used to understand network properties and graph algorithms. Key to such analyses are the different parameters of each model, which affect various network features, such as its size, clustering, or degree distribution. The exact effect of the parameters on these features is not well understood, mainly because we lack tools to thoroughly investigate this relation. Moreover, the parameters cannot be considered in isolation, as changing one affects multiple features.
  Existing approaches for finding the best model parameters of desired features, such as a grid search or estimating the parameter--feature relations, are not well suited, as they are inaccurate or computationally expensive.

  We introduce an efficient iterative fitting method, named \emph{ParFit}, that finds parameters using only a few network samples, based on the Robbins-Monro algorithm. We test ParFit on three well-known graph models, namely Erdős--Rényi, Chung--Lu, and geometric inhomogeneous random graphs, as well as on real-world networks, including web networks. We find that ParFit performs well in terms of quality and running time across most parameter configurations.
\end{abstract}





\maketitle

\section{Introduction}
\label{sec:parameter_fitting_introduction}

In the current age of big data, a lot of important information exists as huge networks, such as interaction networks, semantic networks, and social networks.
Their immense size makes certain tasks very challenging, for example, sharing, processing, or reasoning about these networks.
A common way to overcome this challenge is to utilize \emph{models}, which aim to describe networks via a small amount of parameters.
Ideally, these model parameters are sufficient to capture all of the important features of real-world networks, thus reducing their complexity and allowing to handle tasks more easily.

A fundamental concept for this purpose are random graph models~\cite{drobyshevskiy2019randomGraphs}.
Given a certain set of parameters, such models do not construct a specific graph but instead a random graph out of a family where most samples are likely to exhibit similar features.
This is oftentimes beneficial, as it, for example, allows to share the structure of a network without sharing the exact network, or to produce different benchmarks with similar properties.
Due to this importance, various random graph models have been analyzed with respect to how well they can reproduce real-world graphs, such as exponential random graph models~\cite{an2016fittingERGMs}, Kronecker graphs~\cite{leskovec2010kroneckerGraphs,gleich2012improvinKroneckerGraphs}, approaches based on clustering~\cite{handcock2007clustering,bansal2009clustering}, approaches based on embeddings~\cite{goyal2018embeddings,boguna2010hyperbolicMapping}, and further models~\cite{gutfraind2015coarseningGraphs,sala2010measurement,leskovecGraphsTimeDensification2005}.
In order for each model to reproduce an input well, it is essential to choose the parameters of the model carefully.
This poses a challenging task and strongly depends on the network models, since the relationship between input parameters and output features might not always be clear.

To this end, suitable parameters are selected in various ways, depending on what information about the model is available.
If applicable, gradient descent~\cite{leskovec2010kroneckerGraphs} or moment-based methods~\cite{gleich2012improvinKroneckerGraphs} are good approaches for finding the log-likelihood maximizer of the parameters.
If no such closed expression is known, an alternative is to estimate parameters based on domain knowledge (e.g., an estimated linear relation, monotonic behavior, or a formula)~\cite{blasius2018towards,nagy2022network}.
In case of a lot of data, techniques from machine learning to train a network distance function can be used in order to choose nearest neighbors of possible parameter configurations~\cite{aliakbary2015noise}.
If (almost) no information is available, running a grid search and choosing the best parameters based on a distance measure~\cite{sala2010measurement,Nagy2019} is a possibility.

Although these approaches can yield good results, they come with their own limitations.
Minimizing the gradient typically results in hard, non-convex optimization problems~\cite{drobyshevskiy2019randomGraphs}.
Similarly, the moment-method usually requires a huge amount of samples, rendering it not very efficient~\cite{gleich2012improvinKroneckerGraphs}.
Domain knowledge needs to be specific to be of use.
For example, the temperature parameter of the geometric inhomogeneous random graph model is known to be negatively correlated to the expected clustering coefficient of the generated networks, but the exact relation is unclear and depends on other model parameters. A linear estimation, as suggested in~\cite{krioukovHyperbolicGeometryComplex2010}, is not very accurate.
Techniques from machine learning usually yield intransparent weight functions and involve a computationally expensive training phase.
Last, a grid search for the best parameters is computationally expensive and has limited accuracy.

Another challenge all approaches have to face is that many random graph models do not guarantee to construct connected graphs, whereas real-world networks in available datasets are often connected.
This is usually not due to the represented real-world network being completely connected but due to the dataset being only its largest connected component, which can originate in the dataset creation method (repeated neighbor selection) or during some preprocessing, since usually only the largest component is of interest.
Determining the largest component of the artificial graph resulting from models might affect some of its features, most notably its number of vertices and its average degree.
Thus, the relation between the input parameters and output features becomes more complex. While there is still a correlation, the exact relation is not known. Previous works mainly deal with models generating connected networks~\cite{Nagy2019}, or estimate the effect on the number of vertices, but not on the other parameters, thus applying the reduction to the largest component after parameter fitting~\cite{blasius2018towards}.

We propose a method, named \emph{ParFit}, that is designed to circumvent these problems.
ParFit fits the parameters of a collection of random network models while requiring only few network model samples to reach well fitting model parameters. This method involves no training phase and can deal with the complex parameter landscape introduced by reducing to the largest component.

\subsection{Setting}
We consider three well-known random graph models, namely, \ER, Chung--Lu with power-law degree distribution, and geometric inhomogeneous random graphs (GIRGs). For all of these models, we consider the variant of the model where only the largest component is returned. For every model, we provide a correspondence of model parameters to measurable network features with a strong correlation; however, changing one parameter can change multiple features, and the effect after only considering the largest component is unclear.
Further, due to the random nature of each model, we only have access to samples from its probability distribution.
Under these constraints, given a model and a graph, the goal is to find model parameters that, in expectation, yield networks with the corresponding features matching those of the given graph.

While there are asymptotic results on the giant components and connectivity of the models based on the parameters~\cite{erdhos1959random,erdosEvolutionRandomGraphs1960,cl-adrgged-02,cl-ccrgg-02,chungVolumeGiantComponent2006,Geome_inhom_rando_graph_jour2019} as well as results on predicting features for the model without reduction to the largest component~\cite{cl-adrgged-02,Geome_inhom_rando_graph_jour2019}, to the best of our knowledge, no formulas for expected values of these models' parameters and features we consider are known, in particular for the GIRG model with reduction to the largest component.
This eliminates the likelihood- and moments-based methods mentioned above.

\subsection{Our Method: ParFit}
We treat our setting as a root-finding problem and present a parameter fitting method for this scenario (\emph{ParFit}, Section~\ref{sec:our-parameter-fitting}) based on the Robbins-Monro method~\cite{robbins1951stochastic} in stochastic approximation. In an iterative process, a model parameter guess is established, and based on a single model network sample with these parameters, the parameters are updated according to the deviation of the sample feature values from the target features values. To account for fluctuations around the optimal parameter values, the final parameters are the mean over the parameters of the most recent iterations.

\subsection{Contribution}
We show the effectiveness of ParFit by evaluating it on a wide range of scenarios, including both random graph models as well as real-world networks. For all three random graph models, we find that ParFit yields parameters that fit very well within only few iterations (Table~\ref{tbl:fitting-accuracy} and Figure~\ref{fig:girg-1d-fit}), even in the difficult regime of low vertex degree, where reducing to the largest component has a large influence on all network features.
For real-world networks (Section~\ref{sec:applications}), we also observe that ParFit works well, including the low-degree regime. Especially, our fitted power-law exponents are similar to those estimated in related work.
Overall, ParFit is effective and only needs few iterations to find suitable model parameters.

\section{Problem Definition}
\label{sec:problem_definition}

Given a network model~$M$ as well as an undirected, unweighted graph $G$, we aim to find
parameters for~$M$ that provide the best fit for $G$ with respect to a well chosen metric.
While the best metric would be to compare the parameters of~$M$ and~$G$, this is generally not possible, as the parameters
of are not directly observable in $G$.
Hence, we choose our metric based on the following observation:
For every parameter, there is usually a corresponding \emph{feature} that is measurable
in any graph, which is mainly controlled by this parameter.  Our goal
is to find model parameters such that these measurable features match
the features of $G$ in expectation.

To formalize this, we call $M$ a \emph{random graph model} with $p$ parameters $\theta = (\theta_1, \dots, \theta_p) \in \mathbb R^p$ if $M(\theta)$ is a probability distribution over the set of all graphs $\mathcal G$, i.e., the outcome of each trial is a graph and $M(\theta)$ assigns a probability to each $G \in \mathcal G$.
We write $H \sim M(\theta)$ to indicate that $H \in \mathcal G$ was sampled from $M(\theta)$.
To improve readability, we add $\theta$ as a subscript to the sampled graph, i.e., $H_\theta \sim M(\theta)$ reminds the reader that $H_\theta$ was sampled using the parameters $\theta$.

A model can be equipped with \emph{measurable features} $\varphi_1, \dots, \varphi_p$, where each measurable feature $\varphi_i$ for $i \in [p]$ is a function mapping a graph to a numerical value, i.e., $\varphi_i\colon \mathcal G \to \mathbb R$.
We assume a one-to-one correspondence between the parameters of the model and the measurable features and say that $\varphi_i$ is the measurable feature \emph{corresponding} to the parameter $\theta_i$ for $i \in [p]$.
A measurable feature $\varphi_i$ is a good choice if it can be efficiently evaluated, if $\varphi_i$ is strongly correlated with its corresponding parameter $\theta_i$, and if it has only a minor dependence on other parameters $\theta_j$ with $j \neq i$.
For brevity, we also write $\varphi(G) = (\varphi_1(G), \dots, \varphi_p(G))$ for a graph $G \in \mathcal G$.

With this formalization, we define the \emph{parameter fitting problem}.
Given a random graph model $M$ equipped with measurable features $\varphi$ and given a graph $G$, find parameter values $\theta$ such that for a graph $H_\theta \sim M(\theta)$ sampled from $M$, the expected measurable features are close to the features of $G$, i.e., $\norm{\E{\varphi(H_\theta)} - \varphi(G)}$ is minimized.

This is essentially a multivariate root-finding problem
(when assuming there is a solution with error $0$), with the difficulty
that the measurable features of sampled graphs might have high
variance.  Thus, we can view this as a stochastic optimization problem
with noisy evaluations.  Moreover, note that the input graph $G$ is
not really used except for evaluating the measurable features.  Thus,
one can also view $\varphi(G)$ as being the input instead of $G$
itself.

\section{Network Models}
\label{sec:network_models}

We consider several random graph models.
Here, we discuss their parameters and our choice of corresponding measurable features.
\paragraph{\ER Graphs.}

Given two parameters $n$ and $k$, the \ER model~\cite{erdhos1959random} generates a graph with $n$ vertices where any pair of vertices is connected with probability $p = k / (n-1)$ independently of all other choices. We consider the model variant that reduces the resulting graph to its largest connected component.

As corresponding measurable features, we use the \emph{number of vertices} of a graph to correspond to the parameter $n$ and the \emph{average degree} to correspond to $k$.  These are canonical choices, as the initially generated graph has $n$ vertices and expected average degree $k$.  Note, however, that only considering the largest connected component makes this connection less direct.
\paragraph{Chung--Lu Graphs.}

In the Chung--Lu model~\cite{aiello2000random,cl-ccrgg-02,cl-adrgged-02}, edges are drawn independently with varying probabilities, allowing for a heterogeneous degree distribution.
Every vertex has a weight, and every pair of vertices is connected with probability proportional to the product of their weights.
The expected degree of each vertex then roughly follows its weight.
We consider a power-law version of this model where the weight distribution follows a power law with exponent $\beta$.
Formally, the model has three parameters $n$, $k$ and $\beta$.
We use $n$ vertices, with vertex $i$ having weight $w_i = c\cdot i^{-1 / (\beta-1)}$, where $c$ is such that the average weight is $k$, i.e., the total weight is $W = kn$.
Then, vertices $u$ and $v$ are adjacent with probability
\begin{equation*}
  p(u, v) = \min(1, w_u w_v / W).
\end{equation*}
We reduce the resulting graph to its largest connected component.

As before, we use the \emph{number of vertices} of a graph as measurable feature corresponding to the parameter $n$ and the \emph{average degree} as measurable feature corresponding to $k$.
The power-law exponent $\beta$ controls the variance of the degree distribution, with lower $\beta$ yielding a higher variance.
A way of measuring this is the \emph{heterogeneity} as defined by \citet{blasiusExternalValidityAverageCase2022}, which is the base 10 logarithm of the coefficient of variation of a graph's degree distribution.
As the heterogeneity is negatively correlated with $\beta$, we use its negation as measurable feature corresponding to $\beta$.

\paragraph{Geometric Inhomogeneous Random Graphs.}

The model of geometric inhomogeneous random graphs (GIRGs)~\cite{Geome_inhom_rando_graph_jour2019} is similar to the Chung--Lu model but adds dependencies between edges, using an underlying geometry.
We use a 1-dimensional geometry for which the GIRG model is closely related to hyperbolic random graphs~\cite{krioukovHyperbolicGeometryComplex2010}.

As in the Chung--Lu model, we have the parameters $n$, $k$, and $\beta$ with similar meanings as before; see details below.
Additionally, there is the \emph{temperature} $T$ as fourth parameter, controlling the strength of the geometry and thereby the amount of dependence between the edges.
Given these parameters, a GIRG is generated as follows.
Start with $n$ vertices and assign each vertex $v$ a random weight $w_v$ following a power-law distribution with exponent $\beta$ and a random position $x_v$ drawn uniformly from the interval $[0, 1]$.
The distance between two vertices $u$ and $v$ with positions $x_u$ and $x_v$ is $\norm{x_u - x_v} = \min(|x_u - x_v|, 1 - |x_u - x_v|)$.
With this, any pair of vertices $u$ and $v$ have an edge between them with probability
\begin{equation*}
p(u, v) = \min\left(1, c{\bigl(w_u w_v / (\norm{x_u - x_v} W)\bigr)}^{1/T}\right),
\end{equation*}
where $W$ is the sum of all weights and the constant $c$ is chosen such that the resulting graph has expected average degree $k$~\cite{Effic_gener_geome_inhom_jour2022}.
Finally, the resulting graph is reduced to its largest connected component.

Like in the Chung--Lu model, as measurable features, we use the \emph{number of vertices} corresponding to the parameter $n$, the \emph{average degree} corresponding to $k$, and the negative \emph{heterogeneity}~\cite{blasiusExternalValidityAverageCase2022} corresponding to $\beta$.
The temperature $T$ affects the influence of the geometry.
A stronger geometry leads to the emergence of more triangles, and a common way to measure the amount of triangles is the \emph{average local clustering coefficient} (\emph{clustering coefficient} for short).
The clustering coefficient is the probability that a random vertex together with two random neighbors form a triangle.
As the clustering coefficient is negatively correlated with the temperature $T$, we use its negation as measurable feature corresponding to~$T$.

\section{Our Parameter Fitting Method}
\label{sec:our-parameter-fitting}

\begin{algorithm}[t]
	\caption{\label{alg:fitting-method}
		Our parameter fitting method \emph{ParFit} that returns for a given random graph model~$M$, measurable features~$\varphi$ and target values~$\varphi(G)$ a set of parameters~$\theta$ such that~$M$ parameterized by~$\theta$ exhibits features close to $\varphi(G)$.	See Section~\ref{sec:our-parameter-fitting} for more details.}
	\KwIn{Model $M$, features $\varphi$, target feature values $\varphi(G)$}
	\KwOut{Model parameters}
	$\theta_0 \gets$ initial configuration\;
	$i_\pm$ $\gets 30$\;
	\For{$i \gets 0$ \KwTo $\infty$}{%
		$H_{\theta_i} \gets$ sample from $M_{\theta_i}$\;
		$\Delta_i \gets \varphi(G) - \varphi(H_{\theta_i})$\;
		$\theta_{i+1} \gets \theta_i + \Delta_i$\;
		\If{\emph{sign change in all features at least once}}{%
			$i_\pm \gets \min(i_\pm, i)$\;
		}
		\If{$i \geq i_\pm$}{%
			$\overline{\theta}_{i} \gets \frac{1}{i - i_\pm + 1} \sum_{j = i_\pm}^{k} \theta_j$\;
			\lIf{\emph{$i \geq i_\pm + 200$ \textbf{or} $\overline{\theta}_{i}$ converged}}{%
				\Return{$\overline{\theta}_{i}$}%
			}
		}
	}
\end{algorithm}

Our method \emph{ParFit} (see Algorithm~\ref{alg:fitting-method}) is an iterative stochastic approximation method.
It is based on the \emph{Robbins-Monro} algorithm~\cite{robbins1951stochastic}, originally introduced to solve a root-finding problem by exploiting the monotonicity of the root function and modifying the result based on whether the current guess is too high or too low.
Similarly, ParFit maintains a choice of parameter values for the parameter fitting problem and iteratively adjusts each parameter randomly based on its corresponding measurable feature.
This results in an anytime algorithm, i.e., it can be interrupted at any time to output its current solution. Thus, when to stop the algorithm is a trade-off between solution quality and run time.

As the standard Robbins-Monro algorithm leads to oscillations around optima, we improve the convergence behavior of ParFit by adding \emph{iterate averaging}~\cite{Ruppert1991,Polyak1992}.
That is, starting from some fixed iteration, we log all subsequent results, and once ParFit is stopped, it returns the arithmetic mean of all logged values as the final solution.
We choose this iteration during the run, based on a extension of an adaptive approach~\cite{kesten1958accelerated} to the multivariate case~\cite{delyon1993accelerated}.

\paragraph{Algorithm Description.}
Given a random graph model~$M$ with measurable features $\varphi$ as well as target values $\varphi(G)$, ParFit starts with an initial parameter choice $\theta_0$ (discussed below). In each iteration~$i$, we compute a single sample $H_{\theta_i} \sim M(\theta_i)$ of the model $M$ with parameters $\theta_i$. Then we compute the new solution $\theta_{i+1}$ from $\theta_i$ by adjusting each input parameter proportional to the deviation $\Delta_i \coloneqq \varphi(G) - \varphi(H_{\theta_i})$ of the corresponding feature of the sampled graph from the target value~$\varphi(G)$. In step~$i$, we scale~$\Delta_i$ by a weight $a_i$, called the \emph{gain}. The dependence of~$a_i$ on~$i$ makes it possible to, e.g., introduce a cool-down by gradually reducing $a_i$, which can prevent oscillation. Formally, we set $\theta_{i+1} = \theta_i + a_i \Delta_i$.

For every feature, we keep track of the earliest iteration in which the sign of~$\Delta_{i}$ is different than that of~$\Delta_{i+1}$. We set~$i_\pm$ to be the earliest iteration in which this has happened for all features, and we store solutions from that point on.

When the algorithm is terminated in iteration~$i^*$, we return the arithmetic mean of all stored values, i.e., $\overline{\theta}_{i^*} \coloneqq \frac{1}{i^* - i_\pm + 1} \sum_{j = i_\pm}^{i^*} \theta_j$.

\paragraph{Parameter Choices.}
A common choice for~$a_i$ is ${(i+1)}^{-\alpha}$ for some non-negative $\alpha$. However, in many contexts, $\alpha=0$ and thus $a_i = 1$ is chosen~\cite{kuan1991convergence,spall2005introduction}, even though this does not guarantee convergence~\cite{spall2005introduction}. In our scenario, this choice yields a good fit (Section~\ref{sec:algorithm-configuration}).

For all models, we configure the algorithm as follows.
For the initial solution~$\theta_0$, we take the number of vertices~$n$ and the average degree to be those of the target values~$\varphi(G)$, and choose an initial temperature of~$0.5$ and an initial power-law exponent of~$3.0$.
Further, we set~$i_\pm$ to be at most~$30$.
Last, we terminate the procedure once the final solution\footnote{That is, the arithmetic mean of all so-far stored solutions.}~$\overline{\theta}_i$ fulfills the following convergence criterion: For the last~$10$ iterations, the relative change of~$\overline \theta_i$ across all parameters is below~$1\,\%$; these thresholds are further analyzed in Section~\ref{sec:algorithm-configuration}. We set a maximum of $200$ such averaging iterations to ensure termination; however, this maximum was never reached in our experiments for our choice of algorithm configuration.

\paragraph{Remarks.}
While in general, one might have to add a factor to $\Delta_i$ to account for imbalanced scaling between parameters and features, this was not necessary for our parameters and features.

We compute $\theta_{i+1}$ via only a single sample $H_{\theta_i}$ from the model, which gives a very coarse estimation of the expected feature values $E[\varphi(H_{\theta_i})]$. However, this is not a problem as repeated iterations mitigate these noisy evaluations. Moreover, the iterate averaging helps reduce the effect of outliers when close to the optimum.

\section{Evaluation}
\label{sec:evaluation}

\begin{table*}[t]
	\caption{For the three considered models and the range of settings described in the setup (Section~\ref{sec:setup}), we provide statistics on the quality of ParFit (Algorithm~\ref{alg:fitting-method}). For the four considered features (number of vertices, average degree, heterogeneity and clustering), we consider the mean actual feature values in relation to the target feature values, and provide the Pearson correlation coefficient as well as the mean absolute error (MAE). Values are left blank if not applicable for the respective models. In addition, we provide the mean number of iterations. Please refer to Section~\ref{sec:randomGraphsResults} for a discussion.}
	\label{tbl:fitting-accuracy}
	\begin{center}
		\begin{tabular}{lrrrrrrrrr}
			\toprule
			Model & \multicolumn{2}{c}{Number of vertices}{} & \multicolumn{2}{c}{Average degree}{} & \multicolumn{2}{c}{Heterogeneity}{}& \multicolumn{2}{c}{Clustering}{} & Iterations \\
			& Pearson & MAE & Pearson & MAE & Pearson & MAE & Pearson & MAE  & \\
			\midrule
			\ER & 0.999 & 2.6 & 0.999 & 0.02 & & & & & 13.6 \\
			Chung--Lu & 0.999 & 6.5 & 0.999 & 0.01 & 0.999 & 0.01 & & & 23.3 \\
			GIRG & 0.999 & 51.1 & 0.999 & 0.02 & 0.998 & 0.02 & 0.999 & 0.004 & 32.2 \\
			\bottomrule
		\end{tabular}
	\end{center}
\end{table*}

We evaluate how well ParFit (Algorithm~\ref{alg:fitting-method}) is able to recover the model parameters of random graph models.
For an evaluation on real-world networks, please refer to Section~\ref{sec:applications}.

We consider a \emph{predictive simulation} in which we fit a model to given networks, take samples based on the fitted parameters, and compare the features of the samples to those of the original networks. In order to only measure the quality of ParFit and not of the model, we choose only networks that actually come from the same network model, ensuring that the fitting is actually possible. We consider different quality measures and discuss the results.

\subsection{Setup}
\label{sec:setup}
We implemented ParFit in Python, using several libraries. For network property analysis as well as the \ER and Chung--Lu model, we utilize the networKit library~\cite{Angriman2022}. For the GIRG model, we employ the efficient generator by \citet{Effic_gener_geome_inhom_jour2022}. The experiments were run on a Macbook Pro with an Apple M1 chip and 16 GB RAM. All code and data is published at \url{https://github.com/PFischbeck/parameter-fitting-experiments}.

We consider a range of parameter configurations for all network models. In particular, for the GIRG model, we choose number of vertices $n=\num{10000}$, average degree ranging from $2$ to $10$, power-law exponent ranging from $2.1$ to $25.0$, and temperature ranging from $0.01$ to $0.9999$. For the \ER model and Chung--Lu model, we let the number of vertices range between \num{1000} and \num{10000}. In total, there are 171, 500 and 500 parameter configurations for the \ER, Chung--Lu and GIRG model respectively.

Recall that ParFit aims to minimize the difference between the target feature values and the expected actual feature values at the fitted parameters.
We aim to measure the quality of ParFit with respect to predicting parameters for scenarios that are actually achievable with the given model, in expectation; we focus on the quality for individual networks in Section~\ref{sec:applications}.
To ensure the target feature values are achievable in expectation, for every considered parameter configuration, we take 50 samples from the model, and consider the mean corresponding features across the samples as the input $\varphi(G)$ for ParFit.
For every input, we run ParFit and take 50 samples based on the fitted parameters. We measure the features of those samples and compare their mean to the feature values given to ParFit.
While our problem definition states that we consider the vector length of those differences, it is more useful to look at each feature individually.
For every feature, we measure the Pearson correlation coefficient as well as the mean absolute error, i.e., the mean of the absolute difference between the value of the target feature value and the mean of the 50 samples across all settings.

\subsection{Results \& Discussion}
\label{sec:randomGraphsResults}

\begin{figure*}[t]
	\centering
	\begin{subfigure}{0.45\textwidth}
		\includegraphics{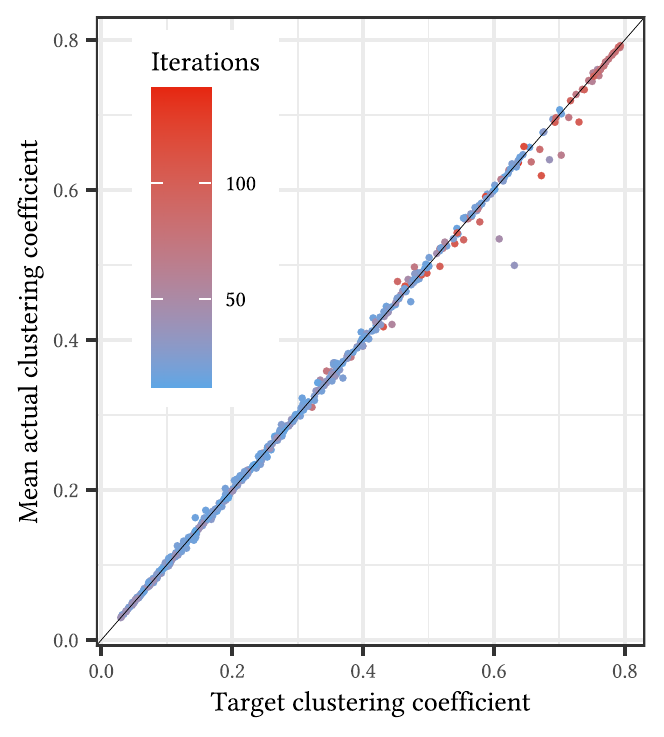}
	\end{subfigure}
	\begin{subfigure}{0.45\textwidth}
		\includegraphics{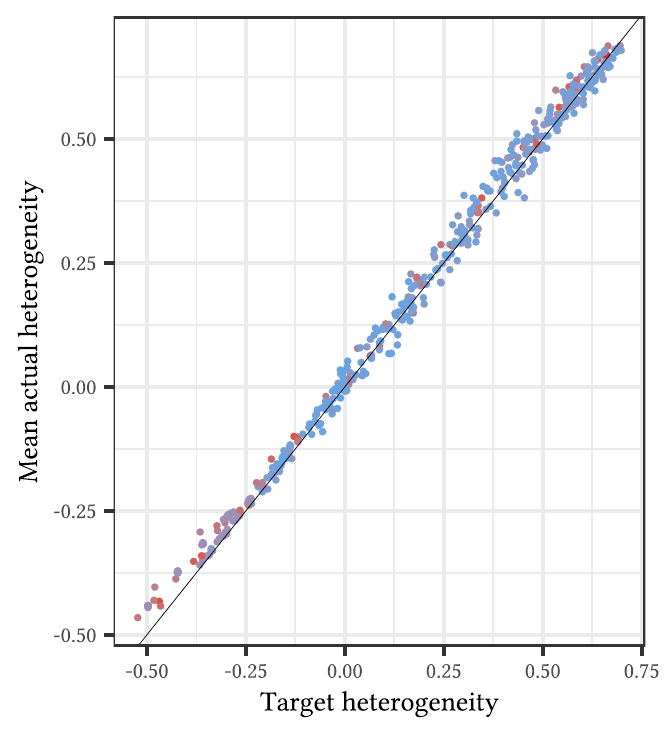}
	\end{subfigure}
	\begin{subfigure}{0.45\textwidth}
		\includegraphics{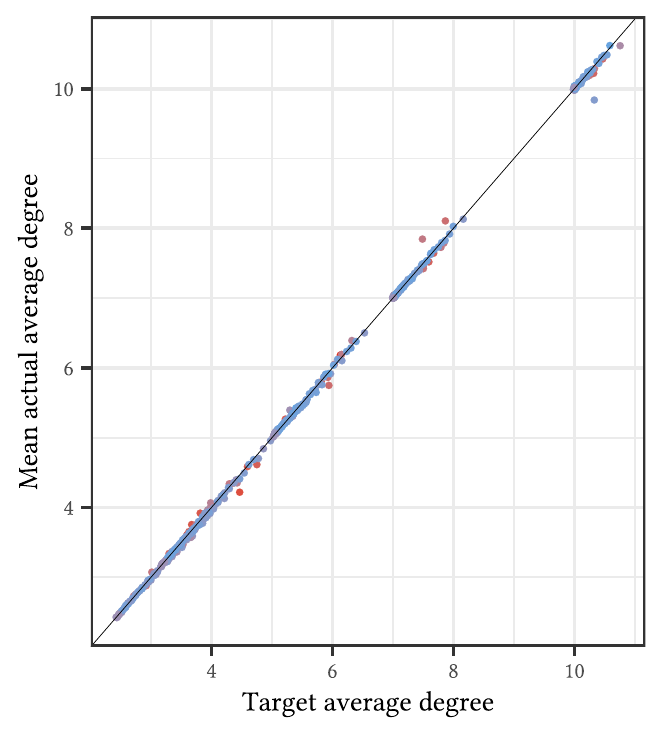}
	\end{subfigure}
	\begin{subfigure}{0.45\textwidth}
		\includegraphics{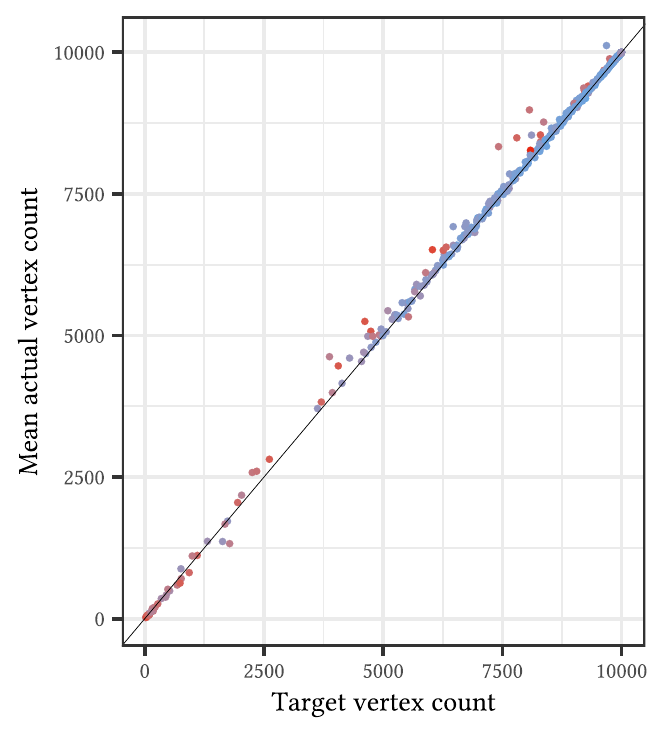}
	\end{subfigure}
	\caption{ParFit (Algorithm~\ref{alg:fitting-method}) evaluated on several scenarios of geometric inhomogeneous random graphs (GIRG; see also Section~\ref{sec:network_models}) instances. We sample 50 GIRG instances from a parameter configuration, take their mean corresponding features, run ParFit, and take 50 samples based on the fitted parameters. For the four relevant features of GIRG instances, the plots show the target feature value (given to ParFit; $x$-axis) as well as the mean actual feature value of 50 samples based on the fitted parameters ($y$-axis). The color shows the number of iterations (i.e., number of samples) taken by ParFit. A darker color indicates fewer iterations. The diagonal line shows the identity function. Please refer to Section~\ref{sec:randomGraphsResults} for a discussion.}
	\label{fig:girg-1d-fit}
\end{figure*}

\begin{table}[t]
	\caption{For the GIRG model and varying $\alpha$ values for the ParFit step size, we provide statistics on ParFit's quality. For the four considered features (number of vertices, average degree, heterogeneity, clustering), we consider mean actual feature values in relation to target feature values, and provide the mean absolute error (MAE). We also provide the mean iteration count. Please refer to Section~\ref{sec:algorithm-configuration} for a discussion.}
	\label{tbl:girg-ablation-alpha}
	\begin{center}
		\input{figures_R/tables/girg-1d_ablation_alpha.tex}
	\end{center}
\end{table}

\begin{table}[t]
	\caption{For the GIRG model and varying relative threshold values \emph{(Thrsh.)} for the convergence stopping criterion, we provide statistics on the quality of ParFit. For the four considered features (number of vertices, average degree, heterogeneity and clustering), we consider mean actual feature values in relation to target feature values, and provide the mean absolute error (MAE). In addition, we provide the mean iteration count. Please refer to Section~\ref{sec:algorithm-configuration} for a discussion.}
	\label{tbl:girg-ablation-threshold}
	\begin{center}
		\input{figures_R/tables/girg-1d_ablation_threshold.tex}
	\end{center}
\end{table}

Table~\ref{tbl:fitting-accuracy} provides an overview of the measured qualities of ParFit across the different models and scenarios. Across all features, we see a very strong Pearson correlation.
Considering the number of vertices, the mean absolute error is very low for the \ER model as well as the Chung--Lu model. It is slightly increased for the GIRG model. Note however that all initial parameter configurations for the GIRG model were chosen such that $n=\num{10000}$, while this parameter varies for the other models. The 90th percentile of the absolute error for GIRGs is at $113.2$, indicating the effect of some outliers, which we consider closer in a moment.
The average degree has a very low mean absolute error across all models.
The other two features of heterogeneity and clustering are strongly correlated too. Recalling that the heterogeneity ranges roughly from $-0.5$ to $0.75$, the MAEs of $0.01$ and $0.02$ are very low. The clustering feature values can range from $0$ to $1$, but the MAE for the GIRG model is only at $0.004$.
The mean iteration count is highest for the GIRG model; however, this can be expected as it has the most parameters and corresponding features out of the considered models.

Figure~\ref{fig:girg-1d-fit} shows the target and mean features of the fitted samples for the GIRG model.
Across all four features (heterogeneity, clustering, average degree, number of vertices), the values of the samples from the fitted parameters closely follow the values given to ParFit. Even in those scenarios we consider very difficult, for example those with low average degree leading to a target vertex count considerably below \num{10000}, ParFit manages to find suitable parameters. We see that the number of iterations is usually highest for those scenarios with extreme target values, i.e., high clustering coefficient and low heterogeneity. The cases of low vertex count are not inherently hard because of this feature value, but rather the target vertex count is very low if the average degree is small.

There are some scenarios where ParFit seems to not give perfectly fitting parameters. In particular, for very low heterogeneity, ParFit finds parameters that are slightly too high. In addition, for high heterogeneity, fitted parameters yield graphs with heterogeneity slightly too high or low.
For the clustering, there are some parameter configurations in the high clustering regime where ParFit struggles to yield matching parameters.
And for the number of vertices, there are several scenarios where the number of vertices of the fitted samples is slightly higher than required. We can also see that those are the cases where the number of iterations taken by ParFit is high.
Overall, there are some outliers where ParFit yielded parameters that do not provide perfectly fitting features. However, these cases are rare and coincide with a high number of iterations.

In Section~\ref{sec:eval-extra} (Appendix), we provide the corresponding figures for the \ER and the Chung--Lu power-law model. In these cases, the parameters fit very well again. In the case of the \ER model in particular, low-degree scenarios lead to a higher number of iterations, due to the increased influence of the reduction to the largest component. However, ParFit deals with these cases easily, finding suitable parameters in under 20 iterations.

For the Chung--Lu power-law model, the method behavior is similar to that of the GIRG model scenario. In particular, the number of vertices and the average degree are mostly tightly fitted, except for few outliers where the method also takes substantially more iterations.
For the heterogeneity, we can see that for scenarios with low heterogeneity, the method usually yields parameters that lead to a slightly too high heterogeneity, and it takes up to 60 iterations in such cases.
Overall, ParFit provides fitting parameters across a wide variety of scenarios for all three considered models.

\subsection{Algorithm Configuration}
\label{sec:algorithm-configuration}

Since ParFit can be configured (Section~\ref{sec:our-parameter-fitting}), we discuss our configuration choice and evaluate the effect of varying its parameters.

\paragraph*{Non-constant Gain.} One aspect of ParFit is the choice of the gain function for the step size. We chose a constant gain of $1$; however, in literature, a gain of ${(k+1)}^\alpha$ is commonly used~\cite{spall2005introduction}, where $k$ is the iteration, and $\alpha$ is a non-negative constant.

We consider a range of values for $\alpha$ ranging from $0$ to $1$ and how this affects the number of iterations as well as the quality of the fitted parameters, across the same scenarios described above. In particular, for the same models and scenarios, we run the altered ParFit version and again measure the resulting features.

Table~\ref{tbl:girg-ablation-alpha} shows the resulting mean absolute error (MAE) values for the GIRG model when considering $\alpha$ values of $0.0$, $0.2$, $0.4$, $0.6$, $0.8$, and $1.0$. See Appendix~\ref{sec:eval-extra} for results on other models. We measure only little change in the number of iterations across all models. However, the MAE across the four considered features increases with increasing $\alpha$, supporting our choice of $\alpha = 0$. The exception is a slightly higher MAE for the number of vertices for low $\alpha$. In contrast, the MAE of the heterogeneity on the GIRG model increases from $0.022$ to $0.055$ when going from $\alpha=0$ to $\alpha=1$. The interaction of higher $\alpha$ with iterate averaging was nicely explained by \citet{spall2005introduction}: With decreased gain, the process gets closer to the optimal parameters from one side only, while iterate averaging performs best when the process moves around these optimal parameters.

\paragraph*{Sign Change Phase.} In our algorithm, we configure $i_\pm$ to be the earliest iteration in which all features had a sign change at some point, but we set it to at most~30 to ensure termination. In the experiments discussed in Section~\ref{sec:randomGraphsResults}, this limit was reached for 160 out of the 1171 algorithm runs. In particular, reaching the limit was most commonly due to no sign changes in the heterogeneity in the low heterogeneity regime. In these scenarios, ParFit struggled to reach such low heterogeneity; see Figure~\ref{fig:girg-1d-fit}.

\paragraph*{Convergence Criterion.}

In our method, we stop the algorithm once the relative change in the current solution was below the threshold $1\,\%$ across all parameters for the last 10 iterations. We consider different values for this threshold to observe the effect on number of iterations and solution quality. Note that we still limit the number of averaging iterations to at most 200. This limit was only reached for 107 out of 5855 algorithm runs.

Table~\ref{tbl:girg-ablation-threshold} shows the resulting mean absolute errors (MAE) for the GIRG model (see Appendix~\ref{sec:eval-extra} for results on other models).
As expected, a higher (i.e., more tolerant) threshold for convergence leads to a decreased number of iterations but also to slightly higher MAE values. Since we still enforce at least 10 iterations, even the solutions where the method is stopped early seem to provide generally good values. Our final choice of $0.01$ for the threshold provides a good middle ground between few iterations and low MAE values.

\section{Applications}
\label{sec:applications}

We apply ParFit (Algorithm~\ref{alg:fitting-method}) with the GIRG model to \num{35} undirected real-world networks from the KONECT database~\cite{konect}.
Table~\ref{tbl:konect-full-table} shows the results.
The expected values for the measurable features were obtained by sampling 50 graphs for the fitted parameters.

\begin{table*}[t]
	\caption{Comparison of four different features (\emph{Actual}; number of vertices, average degree, heterogeneity, clustering) of real-world networks \emph{(Graph)} to the average of those features of~$50$ samples from ParFit (Algorithm~\ref{alg:fitting-method}) fitted to each network \emph{(Measured)}, using the GIRG model (Section~\ref{sec:network_models}). The column following each feature shows the respective GIRG model parameter from ParFit. The last column is the number of iterations that ParFit ran before terminating. See Section~\ref{sec:applications} for a discussion.}
	\label{tbl:konect-full-table}
	\begin{center}
		\input{figures_R/tables/girg-1d_real-world.tex}
	\end{center}
\end{table*}

\paragraph{Quality of Fit.}

We first discuss the number of vertices and the average degree.
For many networks, the average degree is sufficiently high such that the model generates a connected graph.
In these cases, simply setting $n$ to the number of vertices yields the correct result.
Similarly, setting the parameter $k$ to the desired average degree works well as the fitter for the average degree of the GIRG generator is already very good~\cite{Effic_gener_geome_inhom_jour2022}.
Although ParFit does not know this, it easily finds the correct values for $n$ and $k$ in these cases.

In the more interesting cases of low average degrees, ParFit works very well.
For all but two networks, the deviation between the desired and the mean number of vertices is below \SI{1}{\percent}.
For the other two, \texttt{Youtube} and \texttt{Hyves}, the error is \SI{1.87}{\percent} and \SI{2.04}{\percent}.
For the average degree, we have a similar picture, with deviations below \SI{1}{\percent}, except for \texttt{Youtube} and \texttt{Hyves} with \SI{2.47}{\percent} and \SI{2.78}{\percent}.
To highlight one example where ParFit provides new capabilities that were not available before, consider the \texttt{Route views} network.
It has \SI{6.5}{k} vertices and an average degree of \num{3.9}.
To generate a connected graph of that size and density with the same clustering and heterogeneity using the GIRG model, it is necessary to generate a graph with \SI{13.4}{k} vertices and average degree \num{1.94}, i.e., twice the number of vertices and half the average degree.

The two other measurable features are less directly controllable by model parameters than the number of vertices and the average degree.
In fact, for some real-world networks, the measurable features might take values that are impossible to produce with the GIRG model.
For the clustering coefficient, the largest deviation between desired and achieved value is only \num{0.06}.
For the heterogeneity, the deviation is also small for most instances.
However, there are some instances with very high or very low heterogeneity, where the deviation is a bit higher.
This is particularly true for the four road networks (\texttt{Roads TX/CA/PA} and \texttt{Chicago roads}).
We want to stress that this is not a flaw of ParFit but rather indicates that the model is not a good representation of these types of networks.
Although GIRGs can be used to generate homogeneous graphs by setting the power-law exponent very high (which is equivalent to giving all vertices the same weight), the degree of each vertex still follows a binomial distribution, which has some amount of variance.
This makes it impossible to generate GIRGs that have the same heterogeneity as regular graphs (or graphs that are too close to being regular).
Detailed figures for the heterogeneity and clustering coefficient can be found in Appendix~\ref{sec:real-world-appendix}.

\begin{figure}[t]
	\centering
	\includegraphics{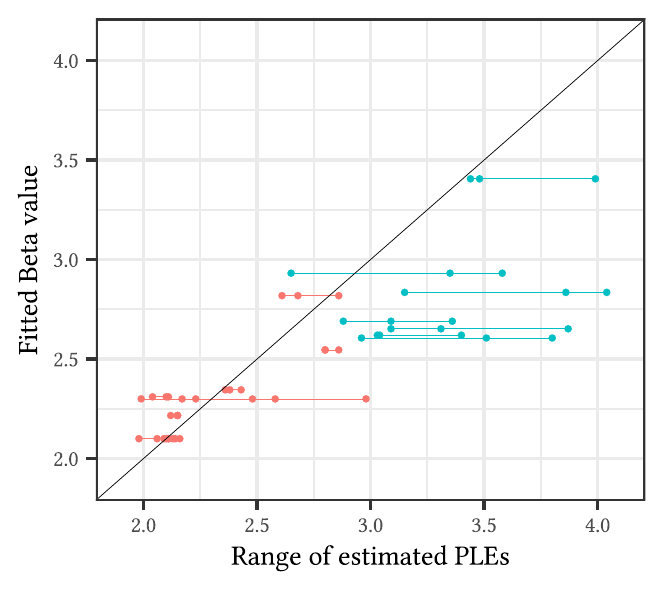}
	\caption{
		The power law exponents (PLEs) of all real-world networks ($x$-axis) versus the respective PLEs fitted by ParFit (Algorithm~\ref{alg:fitting-method}, $y$-axis) assuming a GIRG model (Section~\ref{sec:network_models}).
		The $x$-axis shows the range of the three PLE estimators in~\cite{voitalov2019scale}.
		Red lines refer to strong power laws, blue to the rest.
		The diagonal is the identity function.
		See Section~\ref{sec:applications} for a discussion.
	}
	\label{fig:konect-scalefree-powerlaw-comparison}
\end{figure}

\paragraph{Temperature and Power-Law Exponent.}

Observe that the fitted temperatures are rather high for most networks.
We offer three potential explanations for this.
First, the influence of a latent underlying geometry is rather low.
Second, the one-dimensional geometry of the model does not fully capture the higher-dimensional nature of the networks.
Increasing the dimension has a similar effect on the clustering coefficient as increasing the temperature.
And third, using the clustering coefficient as a measure for the strength of the underlying geometry oversimplifies matters.
An example for this are the road networks, where it is reasonable to assume that the influence of an underlying geometry is rather high.
However, the considered road networks have a low clustering coefficient and thus require high temperature.
This is due to connections being often subdivided, i.e., there are many vertices of degree~2, which heavily decreases the number of triangles.
We believe that all three explanations have merit for different networks and that it is an interesting future question to study which explanation is the right one for which network.
We note that, though this is beyond the scope of this paper, ParFit enables the study of these kind of questions by running experiments on models with higher dimensions or other measurable features corresponding to the temperature.

The values we obtain for the power-law exponents are not surprising.
Figure~\ref{fig:konect-scalefree-powerlaw-comparison} shows them in comparison to the results of \citet{voitalov2019scale}, who applied three different estimation methods.
The figure includes all networks that were classified as power-law networks~\cite{voitalov2019scale}.
In most cases, our obtained power-law exponents are within the range obtained by \citeauthor{voitalov2019scale} or close to it.

This is insofar interesting, as the two approaches have different objectives.
The goal of \citet{voitalov2019scale} is to find an exponent such that the observed tail-distribution best fits a power-law distribution with that exponent, potentially taking cutoffs into account.
ParFit on the other side aims at finding a power-law distribution whose variance best fits the variance observed in the degree distribution.
Both approaches lead to similar results, indicating the heterogeneity measure we use is well suited as a proxy for the power-law exponent.
Additionally, differences between the resulting power-law exponents may in part come from the fact that we reduced the networks to their largest connected component, while \citet{voitalov2019scale} considered the whole network's distribution.

\section{Conclusion}
\label{sec:parameter_fitting_conclusion}
We have presented a fitting method (ParFit; Algorithm~\ref{alg:fitting-method}) that, given an \ER, Chung--Lu or GIRG model as well target feature values, finds model parameters that lead to the desired features on average. We have shown that ParFit works well for a wide range of scenarios, assuming the target feature values are achievable by the model. We have applied ParFit to real-world networks covering several areas, including infrastructure and online social networks. In these cases, ParFit still provides well-fitting parameters and closely matches related work on the scale-freeness of these networks.

We think our work is applicable to other random graph models, including hyperbolic random graphs. Our work now allows for the use of accurate parameter fitting in many contexts, e.g., further fitting the models to real-world networks in order to properly measure how realistic other features of the sampled model instances are. Further improvements to the fitting method are also possible, e.g., a better approximation of the gain via the Kiefer–Wolfowitz algorithm in the simultaneous perturbation variant.

\FloatBarrier

\balance
\bibliographystyle{ACM-Reference-Format}
\bibliography{references}

\clearpage

\appendix

\section{Additional Evaluation Results}
\label{sec:eval-extra}
In this section, we provide further figures for the \ER model and Chung--Lu power-law model that did not fit into the main paper.

\begin{figure*}[t]
	\centering
	\begin{subfigure}{0.45\textwidth}
		\includegraphics{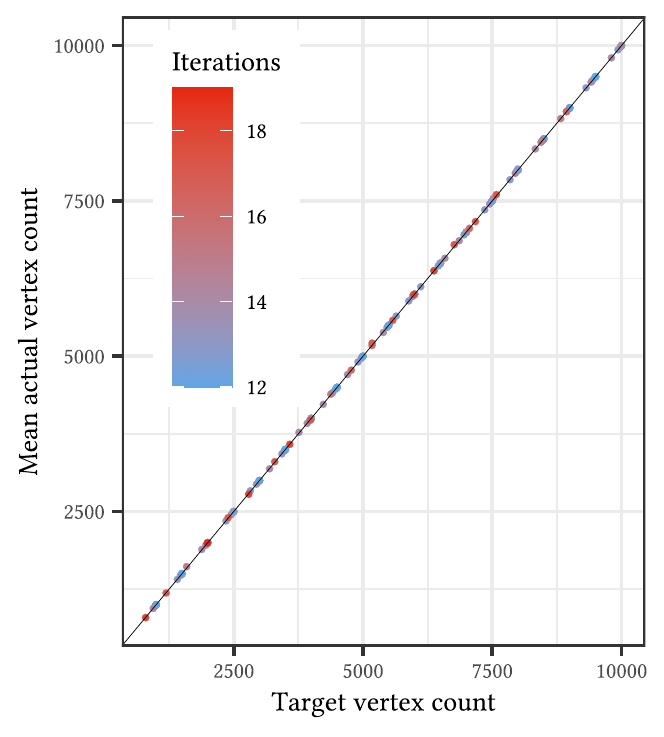}
	\end{subfigure}
	\begin{subfigure}{0.45\textwidth}
		\includegraphics{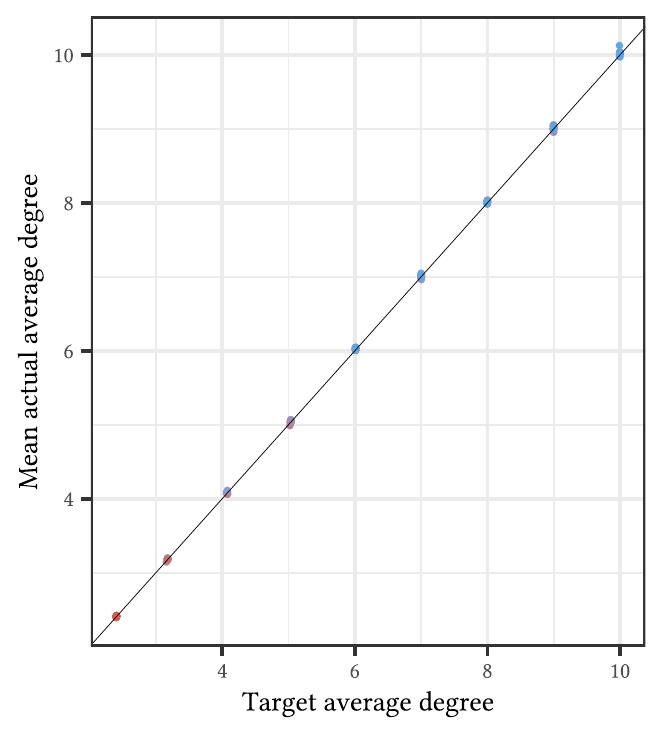}
	\end{subfigure}
	\caption{ParFit (Algorithm~\ref{alg:fitting-method}) evaluated on several \ER scenarios. We sample 50 \ER instances from a parameter configuration, take their mean corresponding features, run the parameter fitting algorithm, and take 50 samples based on the fitted parameters. For the two relevant features of \ER instances, the plots show the target feature value (given to ParFit; $x$-axis) as well as the mean actual feature value of 50 samples based on the fitted parameters ($y$-axis). The color shows the number of iterations (i.e., number of samples) taken by ParFit. A darker color indicates fewer iterations. The diagonal line shows the identity function.}
	\label{fig:erdos-renyi-fit}
\end{figure*}

\begin{figure*}[t]
	\centering
	\begin{subfigure}{0.45\textwidth}
		\includegraphics{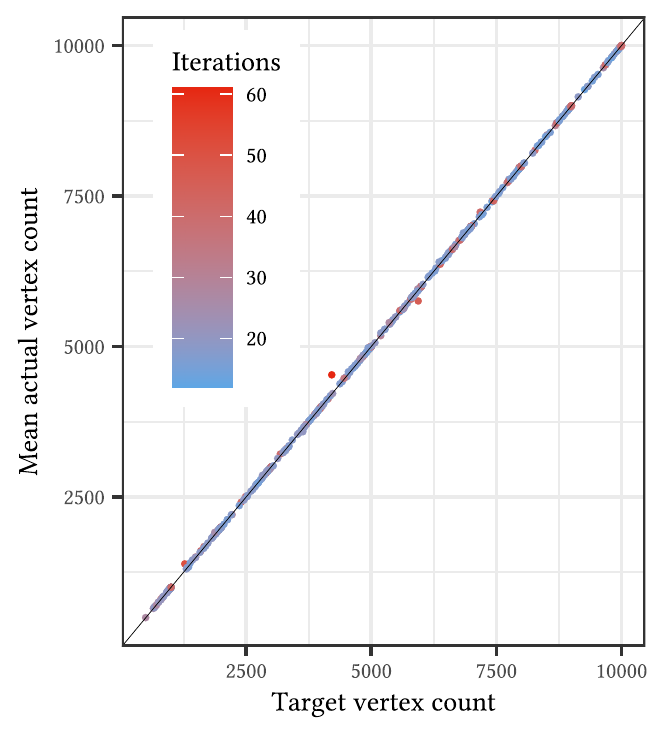}
	\end{subfigure}
	\begin{subfigure}{0.45\textwidth}
		\includegraphics{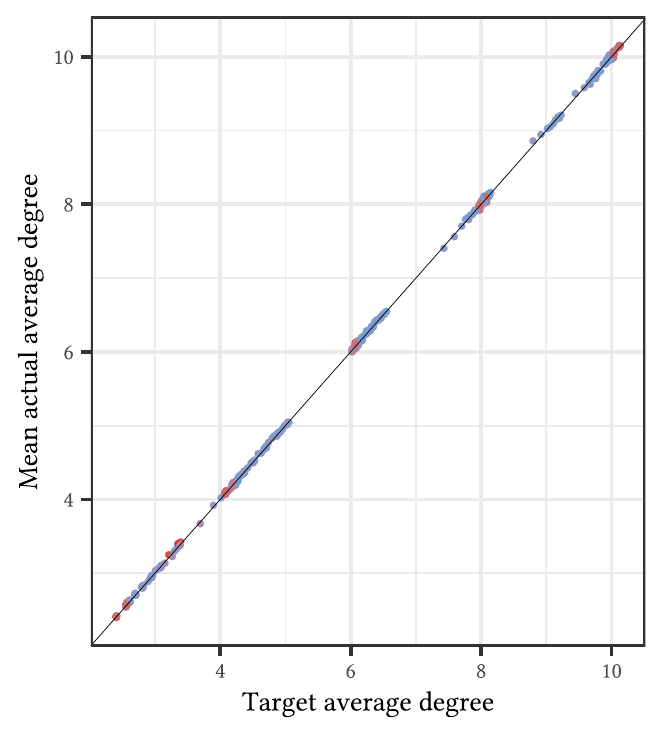}
	\end{subfigure}
	\begin{subfigure}{0.45\textwidth}
		\includegraphics{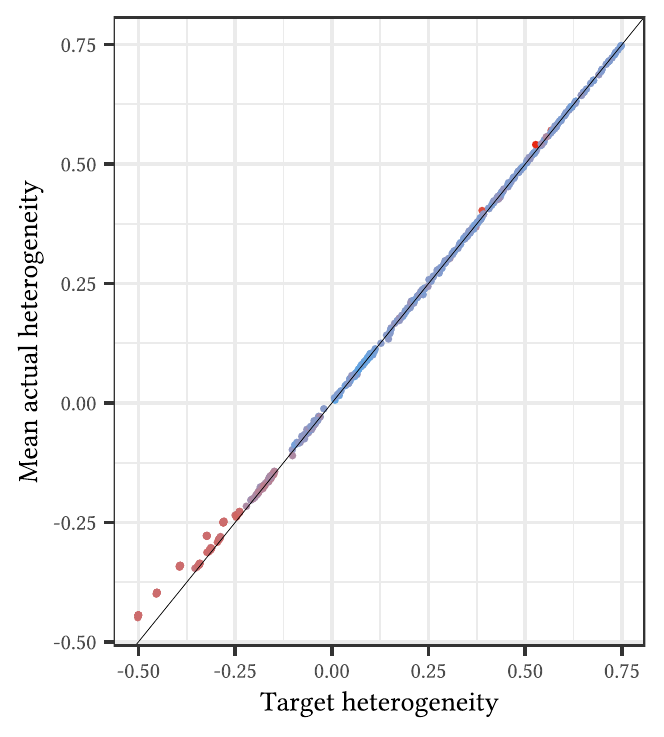}
	\end{subfigure}
	\caption{ParFit (Algorithm~\ref{alg:fitting-method}) evaluated on several Chung--Lu scenarios. We sample 50 Chung--Lu instances from a parameter configuration, take their mean feature values, run the parameter fitting algorithm, and take 50 samples based on the fitted parameters. For the three relevant features of Chung--Lu instances, the plots show the target feature value (given to ParFit; $x$-axis) as well as the mean actual feature value of 50 samples based on the fitted parameters ($y$-axis). The color shows the number of iterations (i.e., number of samples) taken by ParFit. A darker color indicates fewer iterations. The diagonal line shows the identity function.}
	\label{fig:chung-lu-fit}
\end{figure*}

Figure~\ref{fig:erdos-renyi-fit} and Figure~\ref{fig:chung-lu-fit} show the differences between the target features and the mean of the fitted samples, similar to those for the GIRG model shown in Figure~\ref{fig:girg-1d-fit}.

% For some reason this is necessary to make the tables visible at all
% \clearpage

\begin{table}[t]
	\caption{For the Chung--Lu model and varying values of $\alpha$ for the ParFit step size, we provide statistics on the quality of ParFit. For the three considered features (number of vertices, average degree, heterogeneity), we consider the mean actual feature values in relation to the target feature values, and provide the mean absolute error (MAE). In addition, we provide the mean number of iterations.}
	\label{tbl:chung-lu-ablation-alpha}
	\begin{center}
		\input{figures_R/tables/chung-lu-pl_ablation_alpha.tex}
	\end{center}
\end{table}

\begin{table}[t]
	\caption{For the \ER model and varying values of $\alpha$ for the ParFit step size, we provide statistics on the quality of ParFit. For the two considered features (number of vertices, average degree), we consider the mean actual feature values in relation to the target feature values, and provide the mean absolute error (MAE). In addition, we provide the mean number of iterations.}
	\label{tbl:er-ablation-alpha}
	\begin{center}
		\input{figures_R/tables/erdos-renyi_ablation_alpha.tex}
	\end{center}
\end{table}

\begin{table}[t]
	\caption{For the Chung--Lu model and varying values for the convergence threshold, we provide statistics on the quality of ParFit. For the three considered features (number of vertices, average degree, heterogeneity), we consider the mean actual feature values in relation to the target feature values, and provide the mean absolute error (MAE). In addition, we provide the mean number of iterations.}
	\label{tbl:chung-lu-ablation-threshold}
	\begin{center}
		\input{figures_R/tables/chung-lu-pl_ablation_threshold.tex}
	\end{center}
\end{table}

\begin{table}[t]
	\caption{For the \ER model and varying values for the convergence threshold, we provide statistics on the number on the quality of ParFit. For the two considered features (number of vertices, average degree), we consider the mean actual feature values in relation to the target feature values, and provide the mean absolute error (MAE). In addition, we provide the mean number of iterations.}
	\label{tbl:er-ablation-threshold}
	\begin{center}
		\input{figures_R/tables/erdos-renyi_ablation_threshold.tex}
	\end{center}
\end{table}

Table~\ref{tbl:chung-lu-ablation-alpha} and Table~\ref{tbl:er-ablation-alpha} show the quality measures for varying values of ParFit gain $\alpha$, similar to those for the GIRG model shown in Table~\ref{tbl:girg-ablation-alpha}.
In a similar fashion, Table~\ref{tbl:chung-lu-ablation-threshold} and Table~\ref{tbl:er-ablation-threshold} show the quality measures for varying values for the convergence threshold, similar to those for the GIRG model shown in Table~\ref{tbl:girg-ablation-threshold}. Across all tables, we see very similar results in that a low $\alpha$ is a good choice, and a medium threshold is a good trade-off between solution quality and iteration count.

% \clearpage

\section{Real-World Networks}
\label{sec:real-world-appendix}

\begin{figure*}[t]

	\centering
	\begin{subfigure}{0.45\textwidth}
		\includegraphics{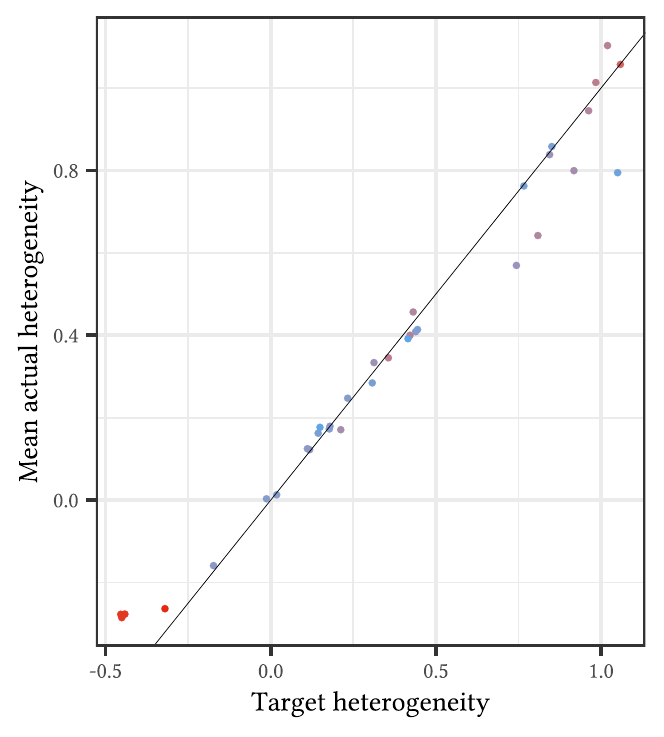}
	\end{subfigure}
	\begin{subfigure}{0.45\textwidth}
		\includegraphics{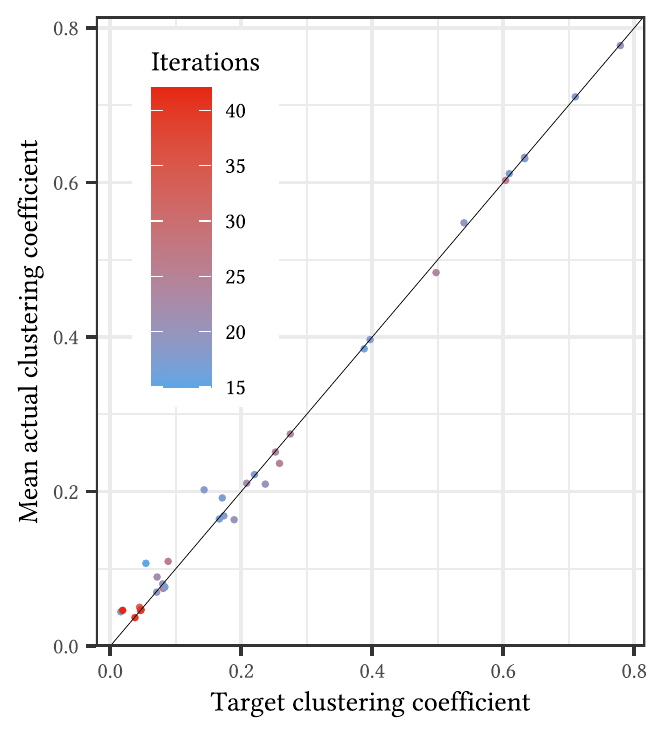}
	\end{subfigure}
	\caption{For every considered real-world network, we run the parameter fitting algorithm for the GIRG model on it, and take 50 samples based on the fitted parameters. The plot shows the true measured heterogeneity/clustering of the real-world network versus the mean heterogeneity/clustering of 50 samples based on the fitted parameters. The color shows the number of iterations (i.e., number of samples) taken by the fitting method. The diagonal line shows an identity function.}
	\label{fig:real-world-girg-1d-het-cc-fit}
\end{figure*}

Figure~\ref{fig:real-world-girg-1d-het-cc-fit} shows information of the fitting method on the real-world networks discussed in Section~\ref{sec:applications} on the GIRG model; in particular on the accuracy in terms of heterogeneity and clustering coefficient.

\end{document}

%% file: figures_R/tables/girg-1d_ablation_alpha.tex
% latex table generated in R 4.3.2 by xtable 1.8-4 package
% Tue Jan 30 15:23:11 2024
\begin{tabular}{rrrrrr}
  \toprule
  $\alpha$ & \multicolumn{4}{c}{Mean absolute error} & Iterations \\
  & Vertices & Avg. deg. & Het. & Clu. & \\
 \midrule
\num{0.0} & \num{50.6} & \num{0.02} & \num{0.02} & \num{0.004} & \num{32.0} \\ 
  \num{0.2} & \num{42.7} & \num{0.02} & \num{0.02} & \num{0.004} & \num{29.0} \\ 
  \num{0.4} & \num{46.9} & \num{0.02} & \num{0.03} & \num{0.004} & \num{28.3} \\ 
  \num{0.6} & \num{63.5} & \num{0.03} & \num{0.04} & \num{0.006} & \num{28.2} \\ 
  \num{0.8} & \num{96.5} & \num{0.04} & \num{0.05} & \num{0.007} & \num{30.3} \\ 
  \num{1.0} & \num{126.9} & \num{0.05} & \num{0.06} & \num{0.009} & \num{32.2} \\ 
   \bottomrule
\end{tabular}

%% file: figures_R/tables/girg-1d_ablation_threshold.tex
% latex table generated in R 4.3.2 by xtable 1.8-4 package
% Tue Jan 30 15:23:17 2024
\begin{tabular}{rrrrrr}
  \toprule
  Thrsh. & \multicolumn{4}{c}{Mean absolute error} & Iterations \\
  & Vertices & Avg. deg. & Het. & Clu. & \\
 \midrule
\num{0.001} & \num{49.9} & \num{0.02} & \num{0.01} & \num{0.003} & \num{104.9} \\ 
  \num{0.005} & \num{48.2} & \num{0.02} & \num{0.02} & \num{0.003} & \num{44.0} \\ 
  \num{0.010} & \num{50.0} & \num{0.02} & \num{0.02} & \num{0.004} & \num{32.4} \\ 
  \num{0.050} & \num{61.1} & \num{0.03} & \num{0.02} & \num{0.004} & \num{21.8} \\ 
  \num{0.100} & \num{61.1} & \num{0.02} & \num{0.02} & \num{0.004} & \num{20.9} \\ 
   \bottomrule
\end{tabular}

%% file: figures_R/tables/girg-1d_real-world.tex
% latex table generated in R 4.3.2 by xtable 1.8-4 package
% Tue Jan 30 15:20:51 2024
\scalebox{0.8}{
\begin{tabular}{lrrrrrrrrrrrrr}
  \toprule
  Graph & \multicolumn{2}{c}{Number of Vertices} & $n$ & \multicolumn{2}{c}{Average degree} & $k$ & \multicolumn{2}{c}{Heterogeneity} & $\beta$ & \multicolumn{2}{c}{Clustering} & $T$ & Iterations \\
  & Actual & Measured & & Actual & Measured & & Actual & Measured & & Actual & Measured & & \\
 \midrule
CAIDA (IN) & \num{26475} & \num{26463} & \num{51917} & \num{4.0} & \num{4.1} & \num{2.1} & \num{0.92} & \num{0.80} & \num{2.10} & \num{0.21} & \num{0.21} & \num{0.79} & \num{22} \\ 
  Skitter (SK) & \num{1694616} & \num{1705244} & \num{1716088} & \num{13.1} & \num{13.0} & \num{12.9} & \num{1.02} & \num{1.10} & \num{2.35} & \num{0.26} & \num{0.24} & \num{0.86} & \num{25} \\ 
  Actor collaborations (CL) & \num{374511} & \num{374511} & \num{374511} & \num{80.2} & \num{80.3} & \num{80.2} & \num{0.31} & \num{0.33} & \num{2.87} & \num{0.78} & \num{0.78} & \num{0.16} & \num{21} \\ 
  Amazon (CA) & \num{334863} & \num{334867} & \num{342937} & \num{5.5} & \num{5.5} & \num{5.4} & \num{0.02} & \num{0.01} & \num{3.41} & \num{0.40} & \num{0.40} & \num{0.59} & \num{18} \\ 
  arXiv (AP) & \num{17903} & \num{17903} & \num{17903} & \num{22.0} & \num{22.0} & \num{22.0} & \num{0.15} & \num{0.18} & \num{2.98} & \num{0.63} & \num{0.63} & \num{0.42} & \num{15} \\ 
  Bible names (MN) & \num{1707} & \num{1710} & \num{1717} & \num{10.6} & \num{10.6} & \num{10.5} & \num{0.23} & \num{0.25} & \num{2.69} & \num{0.71} & \num{0.71} & \num{0.32} & \num{18} \\ 
  Brightkite (BK) & \num{56739} & \num{56792} & \num{57758} & \num{7.5} & \num{7.5} & \num{7.4} & \num{0.44} & \num{0.41} & \num{2.61} & \num{0.17} & \num{0.17} & \num{0.87} & \num{18} \\ 
  Catster (Sc) & \num{148826} & \num{148826} & \num{148826} & \num{73.2} & \num{73.2} & \num{73.2} & \num{1.05} & \num{0.79} & \num{2.10} & \num{0.39} & \num{0.38} & \num{0.92} & \num{16} \\ 
  Catster/Dogster (Scd) & \num{601213} & \num{601213} & \num{601214} & \num{52.1} & \num{52.1} & \num{52.1} & \num{0.96} & \num{0.94} & \num{2.31} & \num{0.50} & \num{0.48} & \num{0.71} & \num{24} \\ 
  Chicago roads (CR) & \num{12979} & \num{12992} & \num{13872} & \num{3.2} & \num{3.2} & \num{3.0} & \num{-0.45} & \num{-0.29} & \num{10.11} & \num{0.04} & \num{0.04} & \num{0.99} & \num{40} \\ 
  DBLP (CD) & \num{317080} & \num{319189} & \num{334410} & \num{6.6} & \num{6.6} & \num{6.4} & \num{0.18} & \num{0.18} & \num{3.10} & \num{0.63} & \num{0.63} & \num{0.36} & \num{19} \\ 
  Dogster (Sd) & \num{426485} & \num{426486} & \num{426488} & \num{40.1} & \num{40.1} & \num{40.1} & \num{0.85} & \num{0.86} & \num{2.22} & \num{0.17} & \num{0.19} & \num{0.98} & \num{17} \\ 
  Douban (DB) & \num{154908} & \num{155044} & \num{174190} & \num{4.2} & \num{4.2} & \num{3.8} & \num{0.44} & \num{0.41} & \num{2.56} & \num{0.02} & \num{0.04} & \num{1.00} & \num{17} \\ 
  U. Rovira I Virgili (A@) & \num{1133} & \num{1133} & \num{1135} & \num{9.6} & \num{9.6} & \num{9.6} & \num{-0.01} & \num{0.00} & \num{2.94} & \num{0.22} & \num{0.22} & \num{0.86} & \num{18} \\ 
  Euro roads (ET) & \num{1039} & \num{1043} & \num{1310} & \num{2.5} & \num{2.5} & \num{2.1} & \num{-0.32} & \num{-0.26} & \num{6.17} & \num{0.02} & \num{0.05} & \num{1.00} & \num{42} \\ 
  Flickr (LF) & \num{1624991} & \num{1625669} & \num{1626051} & \num{19.0} & \num{19.0} & \num{19.0} & \num{0.84} & \num{0.84} & \num{2.47} & \num{0.19} & \num{0.16} & \num{0.89} & \num{20} \\ 
  Flickr (FI) & \num{105722} & \num{105722} & \num{105722} & \num{43.8} & \num{43.8} & \num{43.8} & \num{0.42} & \num{0.40} & \num{2.49} & \num{0.09} & \num{0.11} & \num{1.00} & \num{26} \\ 
  Flixster (FX) & \num{2523386} & \num{2539464} & \num{2649795} & \num{6.3} & \num{6.2} & \num{6.0} & \num{0.77} & \num{0.76} & \num{2.53} & \num{0.08} & \num{0.08} & \num{0.94} & \num{17} \\ 
  Gowalla (GW) & \num{196591} & \num{197260} & \num{198829} & \num{9.7} & \num{9.6} & \num{9.6} & \num{0.74} & \num{0.57} & \num{2.55} & \num{0.24} & \num{0.21} & \num{0.83} & \num{20} \\ 
  Hamsterster (Shf) & \num{1788} & \num{1787} & \num{1789} & \num{14.0} & \num{14.0} & \num{14.0} & \num{0.18} & \num{0.17} & \num{2.46} & \num{0.14} & \num{0.20} & \num{1.00} & \num{18} \\ 
  Hamsterster (Sh) & \num{2000} & \num{2000} & \num{2000} & \num{16.1} & \num{16.2} & \num{16.1} & \num{0.12} & \num{0.12} & \num{2.90} & \num{0.54} & \num{0.55} & \num{0.52} & \num{19} \\ 
  Hyves (HY) & \num{1402673} & \num{1431249} & \num{1983390} & \num{4.0} & \num{3.9} & \num{2.8} & \num{1.06} & \num{1.06} & \num{2.30} & \num{0.04} & \num{0.05} & \num{0.99} & \num{32} \\ 
  LiveJournal (Lj) & \num{5189808} & \num{5189824} & \num{5189986} & \num{18.8} & \num{18.8} & \num{18.8} & \num{0.43} & \num{0.46} & \num{2.84} & \num{0.27} & \num{0.27} & \num{0.75} & \num{24} \\ 
  Livemocha (LM) & \num{104103} & \num{104103} & \num{104103} & \num{42.1} & \num{42.1} & \num{42.1} & \num{0.42} & \num{0.39} & \num{2.50} & \num{0.05} & \num{0.11} & \num{1.00} & \num{15} \\ 
  Orkut (OR) & \num{3072441} & \num{3072441} & \num{3072441} & \num{76.3} & \num{76.3} & \num{76.3} & \num{0.31} & \num{0.28} & \num{2.93} & \num{0.17} & \num{0.16} & \num{0.84} & \num{17} \\ 
  Power grid (UG) & \num{4941} & \num{4942} & \num{6356} & \num{2.7} & \num{2.7} & \num{2.2} & \num{-0.17} & \num{-0.16} & \num{3.85} & \num{0.08} & \num{0.08} & \num{0.88} & \num{19} \\ 
  Proteins (Mp) & \num{1458} & \num{1448} & \num{2303} & \num{2.7} & \num{2.7} & \num{1.8} & \num{0.11} & \num{0.12} & \num{2.65} & \num{0.07} & \num{0.07} & \num{0.96} & \num{19} \\ 
  Reactome (RC) & \num{5973} & \num{5973} & \num{5973} & \num{48.8} & \num{48.8} & \num{48.8} & \num{0.14} & \num{0.16} & \num{2.85} & \num{0.61} & \num{0.61} & \num{0.46} & \num{17} \\ 
  Roads CA (RO) & \num{1957027} & \num{1952063} & \num{2203991} & \num{2.8} & \num{2.8} & \num{2.5} & \num{-0.45} & \num{-0.28} & \num{10.52} & \num{0.05} & \num{0.05} & \num{0.91} & \num{40} \\ 
  Roads PA (RD) & \num{1087562} & \num{1089165} & \num{1222095} & \num{2.8} & \num{2.8} & \num{2.6} & \num{-0.45} & \num{-0.28} & \num{10.37} & \num{0.05} & \num{0.05} & \num{0.91} & \num{40} \\ 
  Roads TX (R1) & \num{1351137} & \num{1355658} & \num{1535134} & \num{2.8} & \num{2.8} & \num{2.5} & \num{-0.44} & \num{-0.28} & \num{10.16} & \num{0.05} & \num{0.05} & \num{0.91} & \num{40} \\ 
  Route views (AS) & \num{6474} & \num{6499} & \num{13432} & \num{3.9} & \num{3.9} & \num{1.9} & \num{0.81} & \num{0.64} & \num{2.10} & \num{0.25} & \num{0.25} & \num{0.70} & \num{24} \\ 
  WordNet (WO) & \num{145145} & \num{145224} & \num{146237} & \num{9.0} & \num{9.0} & \num{9.0} & \num{0.36} & \num{0.35} & \num{2.82} & \num{0.60} & \num{0.60} & \num{0.45} & \num{26} \\ 
  Youtube (CY) & \num{1134890} & \num{1156102} & \num{1339791} & \num{5.3} & \num{5.1} & \num{4.5} & \num{0.98} & \num{1.01} & \num{2.30} & \num{0.08} & \num{0.07} & \num{0.98} & \num{26} \\ 
  Human PPI (MV) & \num{2783} & \num{2790} & \num{3033} & \num{4.3} & \num{4.3} & \num{4.0} & \num{0.21} & \num{0.17} & \num{2.62} & \num{0.07} & \num{0.09} & \num{1.00} & \num{22} \\ 
   \bottomrule
\end{tabular}
}

%% file: figures_R/tables/chung-lu-pl_ablation_alpha.tex
% latex table generated in R 4.3.2 by xtable 1.8-4 package
% Tue Jan 30 15:20:38 2024
\begin{tabular}{rrrrr}
  \toprule
  $\alpha$ & \multicolumn{3}{c}{Mean absolute error} & Iterations \\
  & Vertices & Avg. deg. & Het. & \\
 \midrule
\num{0.0} & \num{6.8} & \num{0.01} & \num{0.01} & \num{23.2} \\ 
  \num{0.2} & \num{5.8} & \num{0.01} & \num{0.01} & \num{24.9} \\ 
  \num{0.4} & \num{6.7} & \num{0.01} & \num{0.02} & \num{27.5} \\ 
  \num{0.6} & \num{10.5} & \num{0.01} & \num{0.02} & \num{31.4} \\ 
  \num{0.8} & \num{21.2} & \num{0.02} & \num{0.03} & \num{36.1} \\ 
  \num{1.0} & \num{44.3} & \num{0.03} & \num{0.05} & \num{37.7} \\ 
   \bottomrule
\end{tabular}

%% file: figures_R/tables/erdos-renyi_ablation_alpha.tex
% latex table generated in R 4.3.2 by xtable 1.8-4 package
% Tue Jan 30 15:21:10 2024
\begin{tabular}{rrrr}
  \toprule
  $\alpha$ & \multicolumn{2}{c}{Mean absolute error} & Iterations \\
  & Vertices & Avg. deg. & \\
 \midrule
\num{0.0} & \num{2.8} & \num{0.01} & \num{13.4} \\ 
  \num{0.2} & \num{2.6} & \num{0.01} & \num{13.6} \\ 
  \num{0.4} & \num{2.5} & \num{0.01} & \num{13.8} \\ 
  \num{0.6} & \num{3.2} & \num{0.01} & \num{14.2} \\ 
  \num{0.8} & \num{4.4} & \num{0.02} & \num{14.8} \\ 
  \num{1.0} & \num{7.4} & \num{0.02} & \num{15.6} \\ 
   \bottomrule
\end{tabular}

%% file: figures_R/tables/chung-lu-pl_ablation_threshold.tex
% latex table generated in R 4.3.2 by xtable 1.8-4 package
% Tue Jan 30 15:20:44 2024
\begin{tabular}{rrrrr}
  \toprule
  Threshold & \multicolumn{3}{c}{Mean absolute error} & Iterations \\
  & Vertices & Avg. deg. & Het. & \\
 \midrule
\num{0.001} & \num{7.6} & \num{0.01} & \num{0.00} & \num{54.5} \\ 
  \num{0.005} & \num{6.1} & \num{0.01} & \num{0.01} & \num{24.4} \\ 
  \num{0.010} & \num{6.7} & \num{0.01} & \num{0.01} & \num{23.3} \\ 
  \num{0.050} & \num{6.7} & \num{0.01} & \num{0.01} & \num{22.7} \\ 
  \num{0.100} & \num{6.6} & \num{0.01} & \num{0.01} & \num{22.8} \\ 
   \bottomrule
\end{tabular}

%% file: figures_R/tables/erdos-renyi_ablation_threshold.tex
% latex table generated in R 4.3.2 by xtable 1.8-4 package
% Tue Jan 30 15:21:16 2024
\begin{tabular}{rrrr}
  \toprule
  Threshold & \multicolumn{2}{c}{Mean absolute error} & Iterations \\
  & Vertices & Avg. deg. & \\
 \midrule
\num{0.001} & \num{2.5} & \num{0.01} & \num{25.8} \\ 
  \num{0.005} & \num{3.2} & \num{0.01} & \num{14.5} \\ 
  \num{0.010} & \num{2.9} & \num{0.01} & \num{13.6} \\ 
  \num{0.050} & \num{3.5} & \num{0.01} & \num{13.3} \\ 
  \num{0.100} & \num{2.8} & \num{0.02} & \num{13.4} \\ 
   \bottomrule
\end{tabular}